\documentclass[%
reprint,
superscriptaddress,
preprintnumbers,
amsmath,
amssymb,
aps,
prb,
floatfix,
twocolumn
]{revtex4-2}

\usepackage{graphicx}% Include figure files
\usepackage{dcolumn}% Align table columns on decimal point
\usepackage{bm}% bold math
\usepackage[utf8]{inputenc}
\usepackage{tabularx}
\usepackage[dvipsnames]{xcolor}
\usepackage{units}
\usepackage[euler]{textgreek}
\usepackage{upgreek}
\usepackage{derivative}
\usepackage{amsmath}
\usepackage{physics}
\usepackage{amssymb}
\usepackage{hyperref}
\hypersetup{colorlinks=true,linkcolor=blue!50!black,urlcolor=blue!50!black,citecolor=blue!50!black}
\usepackage{colortbl}
\usepackage{svg}
\usepackage{romannum}
\usepackage{booktabs}
\usepackage{array}
\usepackage{makecell}
\usepackage[margin=1in]{geometry}  % To control margins
\usepackage{dsfont}
\usepackage{float}
\usepackage[caption=false]{subfig}
\usepackage[justification=justified]{caption}
\usepackage{ragged2e}
\newcommand{\vecperp}{\vec{v}^\perp}

\newcommand{\DMS}{\vec{S}_\chi}
\newcommand{\NUCS}{\vec{S}_N}
\newcommand{\qvec}{\frac{\vec{q}}{m_N}}
\newcommand{\OPER}{\mathcal{O}}
\newcommand{\scat}{\mathcal{M}}

\newcommand\freefootnote[1]{%
  \let\thefootnote\relax%
  \footnotetext{#1}%
  \let\thefootnote\svthefootnote%
}

\begin{document}
\preprint{HIP-2026-21/TH}
\pagenumbering{arabic}
\title{DAMA/LIBRA dark matter signal: a global analysis}

\author{Matti Heikinheimo}
\email{matti.heikinheimo@helsinki.fi}
\affiliation{Department of Physics, P.O.Box 64, FI-00014 University of Helsinki, Finland}
\affiliation{Helsinki Institute of Physics, P.O.Box 64, FI-00014 University of Helsinki, Finland}

\author{Sebastian Sassi}
\email{sebastian.k.sassi@helsinki.fi}
\affiliation{Department of Physics, P.O.Box 64, FI-00014 University of Helsinki, Finland}
\affiliation{Helsinki Institute of Physics, P.O.Box 64, FI-00014 University of Helsinki, Finland}

\author{Alex Stendahl}
\email{alex.stendahl@helsinki.fi}
\affiliation{Department of Physics, P.O.Box 64, FI-00014 University of Helsinki, Finland}
\affiliation{Helsinki Institute of Physics, P.O.Box 64, FI-00014 University of Helsinki, Finland}

\author{Kimmo Tuominen}
\email{kimmo.i.tuominen@helsinki.fi}
\affiliation{Department of Physics,  P.O.Box 64, FI-00014 University of Helsinki, Finland}
\affiliation{Helsinki Institute of Physics, P.O.Box 64, FI-00014 University of Helsinki, Finland}

\author{Niklas Zimmermann}
\email{niklas.zimmermann@helsinki.fi}
\affiliation{Department of Physics, P.O.Box 64, FI-00014 University of Helsinki, Finland}
\affiliation{Helsinki Institute of Physics, P.O.Box 64, FI-00014 University of Helsinki, Finland}

%\date{\today}

\begin{abstract}
    The annual modulation reported by the DAMA/LIBRA collaboration remains in tension with the null results of other direct-detection experiments. We investigate the compatibility of DAMA/LIBRA with LZ, PICO-60 and CDMSlite within a generalized non-relativistic effective field theory (EFT) framework. In addition to the EFT operator expansion we investigate additional modifications to the standard WIMP picture, including inelastic scattering, variations in the NaI quenching factors, and a rescaling of the dark-matter velocity dispersion. We perform a Markov chain Monte Carlo analysis in this extended model space, using two alternative approaches to the likelihood function describing the DAMA/LIBRA data. Both approaches result in strong discrepancy between the data from DAMA/LIBRA and other experiments. We conclude that the extended model space considered here cannot accommodate the DAMA/LIBRA signal while remaining consistent with the non-observation of dark matter in other experiments.

\end{abstract}

\maketitle
%%%%%%%%%%%%%%%%%%%%%%%%%%%%%%%%%%%%%%%%%%%%%%%
%%%%%%%%%%%%%%%%%%%%%%%%%%%%%%%%%%%%%%%%%%%%%%%
\section{Introduction}
\label{introduction}
The paradigm of cold dark matter (DM) remains one of the most intriguing challenges of modern particle cosmology. Various observations of galaxies, galaxy clusters and the cosmic microwave background favour the existence of DM \cite{clowe_direct_2006, collaboration_planck_2020, calabrese_atacama_2025, harvey_non-gravitational_2015, sofue_rotation_2001}. Assuming DM to be a new elementary particle, described in a framework extending the Standard Model of elementary particles and their interactions, implies that DM and ordinary matter are somehow connected. This assumption has sparked the search for DM particles for decades. The most prominent proposed category of models for DM is the Weakly Interacting Massive Particle (WIMP), which should be possible to detect via its interactions with ordinary atoms. Direct detection experiments aim to detect DM particles scattering elastically off nuclei in the target material. The mass range of such WIMPs spans from GeV to TeV inducing nuclear recoil signals within the keV energy scale.

The search for WIMPs has brought forward many null results, imposing strong constraints on the interaction between DM and nucleons \cite{xenonnt_2025, lz_2024, pandax_2025, darkside-50_2024, edelweiss_2020, NEWS_G_2018, CDMSlite_2018, PICO_2019, cosine-100_2024, anais-112_2024}. Alongside the null results, there have been reports of possible DM signals from DAMA/LIBRA \cite{dama_libra_2025}, CRESST-II \cite{cresst_2012}, CoGeNT \cite{cogent_2014} and CDMS II \cite{cdms_si_2013}. A DM origin for the latter three were ruled out by misunderstood background events giving rise to false DM signals \cite{cresst_results_2014, davis_2014}. This leaves DAMA/LIBRA as the only experiment claiming to have measured a positive signal, supported by a fundamental statistical significance $>13\sigma$ for the presence of an annual modulation component in the event rate. However, other sodium iodine (NaI) experiments, such as ANAIS-112 and COSINE-100, have not observed annual modulation. The combined data sets of ANAIS-112 and COSINE-100 indicate a discrepancy of $4.7\sigma$ with the DAMA/LIBRA annual modulation signal \cite{combined_2025}, although this is subject to some uncertainty in the relative energy calibration between the experiments \cite{Busoni:2025tqa}.

Constraints on the DM-nucleon coupling are commonly reported in terms of the spin-independent (SI) constant cross section. The strongest bounds, established by LZ, are of the order $\unit[10^{-48}]{cm^2}$ for a DM mass of $\approx\unit[40]{GeV}$ \cite{lz_2024}. Recently, LZ has reported a single unidentified event at $248 \pm 23 \text{(stat)} \pm 23 \text{(sys)}\,\mathrm{keV}$, though it still awaits confirmation with additional data \cite{LZ:2026axp}. In addition, experiments report the cross section in the spin-dependent (SD) regime \cite{lz_2024, xenonnt_2025, PICO_2019}. Moreover, apart from standard case studies, more advanced analyses using effective field theory (EFT) operator approaches in the non-relativistic limit have been applied to multiple experiments \cite{lzInelastic, xenonnt_effective_2024}.

We note that an EFT analysis has been carried out using COSINE-100 and DAMA/LIBRA data, preserving the possibility that these data sets are compatible \cite{cosine-100_effective_2019}.It should be noted, however, that these analyses only show strong incompatibility after including several experiments with various target materials. The most relevant experiments for achieving such strong discrepancy with the DAMA/LIBRA signal appear to be PICO-60, using ${\rm CF_3I}$ and ${\rm C_3F_8}$ targets and either LUX or XENONnT, using a ${\rm Xe}$ target
\cite{catenaDAMAConfrontsNull2016, kangProbingDAMALIBRA2019}. Introducing inelastic scattering processes cannot remove the tension, as shown in \cite{kangProbingDAMALIBRA2019}. The PICO \cite{PICO:2023uff} and LZ \cite{lzInelastic} collaborations have applied the inelastic scattering formalism to set further constraints. The DAMA/LIBRA collaboration has argued that comparisons between its results and those of other experiments are inconclusive \cite{BERNABEI2020103810}. In particular, they emphasize that detector specific properties, such as the quenching factors and exposure, can significantly affect the inferred compatibility between experiments. This motivates the development of a framework that combines the model-independent non-relativistic EFT formalism with freely varying detector properties and further additional modeling assumptions to account for as many detector specific and systematic uncertainties as possible.

In this work we study the compatibility between DAMA/LIBRA and a selection of other experiments, namely LZ (Xe) \cite{lz_2024}, CMDSlite (Ge) \cite{CDMSlite_2018}, PICO-60 (C$_3$F$_8$) \cite{PICO_2019} and PICO-60 (CF$_3$I) \cite{PICO:2015pux}. We focus on the differences between these diverse experiments using various target nuclei and isotopes, and investigate the possibility of nucleus and isotope dependent signals observed in the DAMA/LIBRA (NaI) experiment. First, we will apply the non-relativistic EFT \cite{fitzpatrick_effective_2013, anand_model-independent_2014} to describe the DM-nucleus interactions. In addition, we apply a modified EFT for inelastic scattering, and allow for variations in the quenching factor and the velocity dispersion in the DM halo model. We perform a global Markov chain Monte Carlo (MCMC) analysis in the theory space spanned by the Wilson coefficients of the EFT and the parameters of the inelastic scattering model, the quenching factors, and the velocity dispersion. The purpose of this analysis is to characterize the level of discrepancy between DAMA/LIBRA and other experiments while keeping the DM model as free as possible, and to observe any correlations between the various parameters in the general theory space that could improve the global fit.

The paper is organized as follows: Section \ref{sec:theory} presents the parameterizations of the models used in our study. Section \ref{sec:global_analysis} gives an overview of our statistical procedure. This is followed by the results in section \ref{sec:results} and conclusions in section~\ref{sec:checkout}.

\section{Theoretical background}
\label{sec:theory}
\subsection{Dark matter velocity distribution}
\label{sec:DMVelDist}

In this work we start from the conventional assumption that the DM halo is fully virialized into an isothermal sphere. This is the Standard Halo Model (SHM). Under this assumption, the DM velocity distribution is fully specified, and given by a Maxwell--Boltzmann distribution cut off at the Milky Way escape velocity $v_\text{esc}$ \cite{savage_compatibility_2009},
\begin{align}
    f\left( \vec{v} \right) = \frac{\exp\left(-\frac{|\vec{v}|^2}{2\sigma_v^2}\right)}{N_{\mathrm{esc}}\sqrt{(2 \pi)^3} \sigma_v^{3}}  \Theta( v_\mathrm{esc} - \left|\vec{v}\right|),
\end{align}
Here $\sigma_v = v_0/\sqrt{2}$ is the velocity dispersion, where $v_0$ is the local circular speed, for which we use the common benchmark value $v_0=\unit[220]{km/s}$. For the escape velocity we use the value $v_\mathrm{esc} = \unit[544]{km/s}$. For an expression of the normalization constant $N_{\mathrm{esc}}$ see, e.g., \cite{savage_compatibility_2009}.

The SHM provides a well-motivated approximation of the DM halo, which is simple to work with since it possesses nice closed form formulae for the dark matter event rate.
However, there is reason to expect the true structure of the DM halo to deviate from the SHM. Apart from the possible existence of finer local structure of the halo, such as streams---see, e.g.,~\cite{kavanagh_new_2015, kavanagh_reconstructing_2016}---observations by the Gaia satellite suggest the existence of a nonvirialized structure in the halo as a result of a past merger event \cite{BelokurovEtAl2018, KruijssenEtAl2018, HelmiEtAl2018}, known as the Gaia sausage. This feature has been modeled with the SHM$^{++}$ velocity distribution \cite{evans_refinement_2019}, which adds an anisotropic component to the SHM velocity distribution to model the sausage. 
A more recent example of a more sophisticated attempt to model the DM halo is the self-consistent Anisotropic Halo Model (scAHM)~\cite{Stanic2025}.

Although the above models would provide more freedom in their parametrization,
alongside a possibility of a more realistic description of the DM halo, they were not computationally feasible for carrying out the present analysis with conventional methods.
However, recently developed methods should make such analyses more viable in the future~\cite{SassiEtAl2025}. In any case, since these models are refinements over the SHM rather than complete departures from it, we expect the SHM to be sufficient for this work and do not believe a more refined description of the velocity distribution would alter the overall conclusions of our study. Therefore, in this work we employ a simple modified version of the SHM, given by
\begin{align}
    f\left( \vec{v} \right) = \frac{\exp\left(-\frac{\vec{v}^2}{2\left(\alpha\sigma_v\right)^2}\right)}{N_{\mathrm{esc}}\sqrt{8 \pi^3 \left(\alpha\sigma_v\right)^{3}}}
      \Theta( v_\mathrm{esc} - \left|\vec{v}\right|),
      \label{eq:halomodel}
\end{align}
where we have taken $\sigma_v \rightarrow \alpha \sigma_v$. The scaling parameter $\alpha$ effectively parametrizes variation of the velocity dispersion from its conventional value. We restrict $\alpha$ to the interval $[0.07, 1]$, where the lower limit is imposed for numerical stability and the upper limit corresponds to the unmodified SHM.

Physically, our modified SHM velocity distributioin may be interpreted as assuming the solar system to be in a colder patch of the halo with a more concentrated velocity distribution.
The motivation of this modified DM halo model is to allow an increase of the annual modulation amplitude of the DM scattering rate, which acts to improve the compatibility between DAMA/LIBRA and other experiments. Thus, this model can act as a simple proxy for characterizing the dependence of the compatibility on the unknown DM halo function.

\subsection{Non-relativistic effective field theory}
\label{sec:NREFT}
Experimental results in DM direct detection are commonly
described as limits on the spin-independent (SI) or the spin-dependent (SD) cross section of DM--nucleus scattering. 
The presence of the latter is dependent on the target nuclei, and therefore the SI cross section is typically used as a basic quantity to compare the constraints across different experiments.

Instead of restricting to the conventional SI/SD description of scattering, in this analysis we employ the general nonrelativistic Effective Field Theory (EFT) description of DM--nucleon interactions introduced in \cite{fan_non-relativistic_2010} and later expanded in \cite{fitzpatrick_effective_2013, anand_model-independent_2014}. This approach provides a complete set of operators expressible in terms of the spins of the target nucleus and the DM particle as well as the recoil momentum and DM velocity. These operators may be ordered in terms of their suppression due to DM velocity and momentum. They include the usual SI and SD interactions at leading order of this expansion. This operator basis enables a systematic study of the potential effects of operators arising from more suppressed interactions.  The EFT operators are \cite{fitzpatrick_effective_2013, anand_model-independent_2014, catena_form_2015}
\allowdisplaybreaks
\begin{align}
    &\OPER_1 = \mathds{1}_\chi \mathds{1}_N \nonumber, \quad
    \OPER_2 =\left( \vecperp \right) ^2 \nonumber, \\
    &\OPER_3 = i \NUCS \cdot \left( \qvec \times \vecperp \right) \nonumber, \quad
    \OPER_4 = \NUCS \cdot \DMS \nonumber,\\
    &\OPER_5 = i\DMS \cdot \left( \qvec \times \vecperp \right) \nonumber, \nonumber \\
    &\OPER_6 = \left( \DMS \cdot \qvec \right) \left( \NUCS \cdot \qvec \right)\nonumber,\\
    &\OPER_7 = \NUCS \cdot \vecperp \nonumber, \quad
    \OPER_8 = \DMS \cdot \vecperp\nonumber,\\
    &\OPER_9 = i\DMS \cdot \left( \NUCS \times \qvec \right)  \nonumber, \quad
    \OPER_{10} = i\NUCS \cdot \qvec  \nonumber,\\
    &\OPER_{11} =i\DMS\cdot\qvec \nonumber, \quad
    \OPER_{12} = \DMS \cdot \left(\NUCS \times \vecperp\right) \nonumber, \quad \\
    &\OPER_{13} = i \left(\DMS \cdot \vecperp\right) \left(\NUCS \cdot \qvec\right) \nonumber, \\
    &\OPER_{14} = i\left(\DMS \cdot \qvec\right) \left(\NUCS \cdot \vecperp\right) \nonumber, \\
    &\OPER_{15} = - \left(\DMS \cdot \qvec\right) \left[\left(\NUCS \times \vecperp\right)\cdot \qvec\right],
     \label{eq:operators}
\end{align}
where $\DMS$ and $\NUCS$ represent the DM and target-nucleon spins, respectively. The relative transverse velocity operator $\vecperp$ is defined such that it is perpendicular to the momentum transfer $\vec{q}$, that is $\vec{q} \cdot \vecperp = 0$, and $m_N$ denotes the average nucleon mass. In this work, we adapt the normalization convention of \cite{anand_model-independent_2014}, resulting in a dimensionless momentum transfer operator and couplings which are generally dimensionless. Furthermore, we work in the isoscalar and isovector basis, and exclude the operator $\OPER_2$ which is not expected to arise from any underlying relativistic theory.

The event rate for DM--nucleus scattering in the EFT is given by
\begin{align}
        \frac{dR}{dE_\mathrm{nr} d\Omega} = &\frac{1}{64 \pi^2} \frac{\rho_\chi}{m_\chi^3 m_T^2} \times \nonumber\\ &\int \langle \left|\scat\right|^2 \rangle \delta(\vec{v}\cdot\hat{q} - v_{\mathrm{min}}) f(\vec{v}) d^3v.
        \label{eq:rateEFT}
\end{align}
Here $\rho_\chi$ is the local DM energy density, $m_\chi$ and $m_T$ are the DM and target nucleus masses, respectively, and $\langle \left|\scat\right|^2 \rangle$ is the squared scattering amplitude averaged over all spins, whose form in the EFT is given in \cite{fitzpatrick_effective_2013, anand_model-independent_2014}. 
The $\delta$-distribution in Eq.~\eqref{eq:rateEFT} arises from energy conservation,
and $v_\text{min} = \sqrt{m_T E_\mathrm{nr} / 2 \mu^2}$ denotes the minimum DM velocity which can produce a scattering event with a given nuclear recoil energy $E_\mathrm{nr}$. 

We classify the operators based on their dependence on the small dimensionless parameters 
$\vec{q}/m_N$ and $\vecperp$.
Hence, we call operators which are independent of both $\vec{q}/m_N$ and $\vecperp$ Leading Order (LO), and operators which are linear in either Next to the Leading Order (NLO). Operators which are quadratic or cubic in any combination of these parameters are similarly labeled N$^2$LO and N$^3$LO, respectively. This categorization of EFT operators is summarized in table \ref{tab:characterization_operators}.
Since SI interaction can be viewed as a special case of SD, we do not treat these cases separately.

\begin{table}
    %\centering
    \begin{tabular}{l c}
        \toprule
        Order & Operators \\
        \midrule
        LO & $\OPER_1 $,  $\OPER_4$ \\
        NLO  & $\OPER_7$, $\OPER_8$, $\OPER_9$, $\OPER_{10}$, $\OPER_{11}$,  $\OPER_{12}$\\
        N$^2$LO & $\OPER_3$, $\OPER_5$, $\OPER_6$, $\OPER_{13}$,  $\OPER_{14}$ \\
        N$^3$LO & $\OPER_{15}$\\
        \bottomrule
    \end{tabular}
    \caption{\justifying The ordering of EFT operators. The momentum and velocity are treated as equivalent expansion parameters to define LO, NLO etc. See text for details.}
     \label{tab:characterization_operators}
\end{table}

We do not cover the theoretically exotic case of the single N$^3$LO operator in our global analysis. Based on our findings for the other operators, the discrepancy of DAMA/LIBRA and other experiments is likely large in this scenario, as there is little room for interference effects that could alleviate the tension in the case of a single operator.

The operators listed in~\eqref{eq:operators} are usually treated in the context of elastic scattering. A model independent treatment of inelastic scattering assumes that the DM particle up-scatters from some initial state $\chi$ to some higher energetic state $\chi^*$~\cite{barello_model_2014}. This results in a change of the DM mass $\delta = m_{\chi^*} - m_\chi$. Furthermore, due to the change in energy, the minimum and transverse velocity need to be redefined as ~\cite{barello_model_2014, bozorgnia_halo-independent_2013}
\begin{align}
    v_\mathrm{min} &= \frac{1}{\sqrt{2m_TE_\mathrm{nr}}} \left(\frac{m_TE_\mathrm{nr}}{\mu} + \delta\right), \label{eq:vmin} \\
    \vecperp_\mathrm{inel} &= \vecperp_\mathrm{el} + \frac{\delta}{\vec{q}^2} \vec{q},
\end{align}
with $\vecperp_\mathrm{el}$ denoting the transverse velocity defined in the elastic scattering processes. We find the inelastic DM nuclear recoil event rate by applying these redefined properties to the operators shown in $\eqref{eq:operators}$ and to the velocity integral in equation \eqref{eq:rateEFT}.

\subsection{Quenching factor}\label{sec:quenching_factor}

DAMA/LIBRA observes the scattering events by measuring scintillation light. The light produced by a nuclear recoil event is quenched compared to an electron recoil with the same energy, while the energy calibration of the detector is based on electron recoil events. Therefore, the ratio of the measured scintillation signals for electron recoils and nuclear recoils must be know in order to translate the observed spectrum into nuclear recoil energy.
This ratio is known as the quenching factor (QF)
\begin{align}
    \text{QF} = \frac{E_{\rm ee}}{E_{\rm nr}}.
\end{align}
The effect of this parameter is that smaller (larger) values of the quenching factor shift the true nuclear recoil spectrum to higher (lower) values of nuclear recoil energy $E_{\rm nr}$ for a given observed energy $E_{\rm ee}$.

Measurements of the ${\rm NaI}$ QF reveal results differing from the DAMA/LIBRA central values, $\text{QF}_\text{Na} = 0.3$ and $\text{QF}_\text{I} = 0.09$~\cite{Bernabei:1996vj, Bernabei:2019ajy}, and also display dependence on energy~\cite{PhysRevC.110.014614,PhysRevC.110.014613,Angloher2025fnl}.
Based on this, we investigate how changes in the assumed DAMA/LIBRA QF impact the compatibility of DAMA/LIBRA and other experiments. We use a parametrization motivated by the Lindhard model~\cite{osti_4701226},
\begin{align}
    \text{QF}(\epsilon) = a \frac{k g(\epsilon)}{1 + k g(\epsilon)},
\end{align}
where $\epsilon(E_{\text{nr}}) = 11.5 E_{\text{nr}}/Z^{7/3}$, with $Z$ the atomic number, and $g(\epsilon)$ is given by
\begin{align}
    g(\epsilon) = 3 \epsilon^{0.15} + 0.7 \epsilon^{0.6} + \epsilon.
\end{align}
The parameter $a$ controls the asymptotic behavior of the QF as $E_{\text{nr}} \gg \unit[1]{keV}$,
while the parameter $k$ controls the slope. Thus in the limit of small $\epsilon$ we have $\text{QF}(\epsilon)=a k \epsilon+{\cal O} (\epsilon^2)$. The limit $k \rightarrow \infty$ yields the constant quenching factor case, $\text{QF}(\epsilon)=a$. 

\section{Global analysis}
\label{sec:global_analysis}

Our global analysis is based on the joint likelihood of measuring the observed data in each experiment. For the LZ, CDMSlite, and PICO-60, the data consists of the number of events passing all cuts, possibly binned in the nuclear recoil energy. Given a model with parameters $\theta$, the likelihood function for a single experiment is given by the Poisson probability
\begin{equation}
        L_\mathrm{P}(\tilde{n}; \theta) = \prod_{j=1}^N \frac{\left(\lambda_j(\theta) + b_j\right)^{n_j}}{n_j!} e^{-(\lambda_j(\theta) + b_j)},
\end{equation}
where $\tilde{n} = \{n_j\}$ is the set of observed numbers of events per bin. The function $\lambda_j(\theta)$ is the expected number of signal events in the $j$-th bin, as given by the model, and $b_j$ is the number of expected background events in the same bin. The joint likelihood for multiple counting experiments is given by the product of their Poisson probabilities.

Instead of reporting the total number of events per bin, the DAMA/LIBRA experiment reports the modulation amplitude of the event rate, binned in nuclear recoil energy. Mathematically, the modulation amplitude can be obtained as the difference between the numbers of events in the time bins corresponding to the maximum and minimum event rate. The corresponding probability distribution is a difference of two Poisson distributions, which is given by the Skellam distribution,
\begin{align}
        \label{eq:skellam}
        &L_\mathrm{S}(\hat{n};\theta) = \prod_{j=1}^N e^{-(\tilde{\lambda}(\theta)_{j, 1} +\tilde{\lambda}(\theta)_{j, 2})} \nonumber \\ &\times \left( \frac{\tilde{\lambda}(\theta)_{j, 1}}{\tilde{\lambda}(\theta)_{j, 2}} \right)^{n_j/2} I_{n_j}\left(2\sqrt{\tilde{\lambda}(\theta)_{j, 1} \tilde{\lambda}(\theta)_{j, 2}}\right).
\end{align}
 Here we use notation combining the expected signal and background events $\tilde{\lambda}_k(\theta) = \lambda_k(\theta) + b_k$. The functions $I_n(z)$ are modified Bessel functions of the first kind. The modulation amplitude data reported by DAMA/LIBRA corresponds to numbers of events per bin on the order of $O(10^4)$. Therefore, we can apply the asymptotic expansion of the modified Bessel function for large arguments \cite{abramowitz+stegun}.
 
 The conversion of modulation amplitudes into Skellam-distributed event counts has the caveat that there is no simple conversion of the reported uncertainties in this method. The likelihood-formula (\ref{eq:skellam}) treats the data as if the uncertainties are dominated by statistical uncertainty, which in the DAMA/LIBRA modulation amplitude seems not to be the case, as the relative uncertainties are of the order of $20\%$, while the expected statistical uncertainty $1/\sqrt{N}$ would be $O(1\%)$ for $O(10^4)$ events per bin. This treatment therefore underestimates the DAMA/LIBRA errors and inflates the importance of the modulation amplitude data in the global fit. For this reason, in the following analysis we will discuss two variations of the global fit: one using the Skellam distribution (\ref{eq:skellam}) for the DAMA/LIBRA likelihood, and another treating the modulation amplitude values as Gaussian distributed with the reported errors. As we will see, the first approach tends to yield best-fit parameters that faithfully reproduce the modulation spectrum in DAMA/LIBRA, and are thus interesting for gauging the level to which the resulting discrepancy with the other experiments can be alleviated while keeping the DAMA/LIBRA signal intact, while the second approach generally yields worse fits to the modulation spectrum but a weaker tension with the other experiments.
 
 To obtain a goodness-of-fit estimation for all experiments, we maximize the joint likelihood function
\begin{align}
\label{eq:EntireLikelihood}
 L(\theta) &= L_\mathrm{P, LZ}(\theta) \times L_\mathrm{P, CDMSlite}(\theta) \nonumber\\
 &\quad\times L_\mathrm{P, PICO}(\theta) \times L_\mathrm{S, DAMA}(\theta).
\end{align}
The parameters $\theta$ contain the DM mass, the EFT Wilson coefficients and the parameters of the QF model, the DM halo model, and the inelastic recoil parameter $\delta$.

The best fit is found by maximizing the likelihood function; that is, by maximizing the probability to draw the observed data from the distribution described by the parameters $\theta$. We apply the standard methodology of minimizing the negative logarithmic likelihood function to find the best fit point.

In order to obtain marginalized posteriors for the parameters we sample the likelihood using a Markov Chain Monte Carlo (MCMC) method. As the proposal function we use the Affine Invariant stretch mover, described in \cite{affine_sampler}. In addition, we use tempering to shorten the initial transient, to avoid local minima, and to enable jumping between degenerate parameters \cite{Swendsen,bioinformatics}. The burn-in period has a time-dependent temperature, where we start all chains in a nonzero tempering that cools down to run levels following an exponential. The temperatures are set using a geometric distribution in the interval $T \in [1, 200]$, where $T = 1$ is the zero degree chain \cite{Rozada_2019}.

To verify that the ensembles converge to the correct minimum, we run the same model multiple times, and ensure that the independent runs achieve the same result. 

\section{Results}
\label{sec:results}

The DAMA/LIBRA data are published as modulation amplitudes binned in recoil energy, as well as residual rates as a signal over time. In this work, we utilize both data sets. We note that correlations between these data sets arise in the fitting procedure, but such correlations cannot be quantified with the publicly available data. Hence we treat the data sets in our analysis as independent. Nevertheless, we find that including both data sets rather than only one leads to only minor changes in the best fit parameters.

In contrast to DAMA, other direct detection experiments---including LZ, CDMSlite, and PICO---have reported no events attributable to dark matter scattering. The only exception is the recently reported single event from LZ matching a 248 keV nuclear recoil \cite{LZ:2026axp}. Since both PICO experiments employ bubble chamber detectors with poor energy resolution, we model them as counting experiments. We adopt the same procedure for LZ, considering the full nuclear recoil search region of interest (ROI) as a single energy bin. Based on the most recent LZ data release, we infer a total of 18 expected background events \cite{lz_2024}.

\begin{figure*}
    \centering
    \subfloat[\label{fig:dama_lz_mod_amp}]{
        \includegraphics[width=0.5\linewidth]{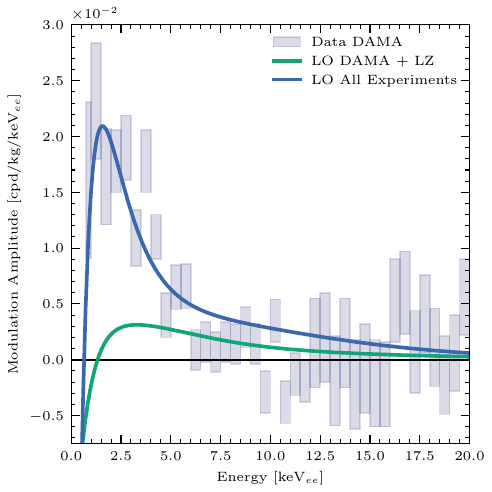}
    }
    \subfloat[\label{fig:dama_lz_posterior}]{
        \includegraphics[width=0.5\linewidth]{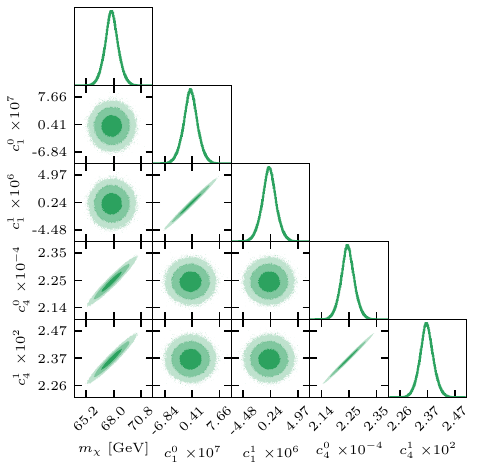}
    }
    \captionsetup{justification=RaggedRight,singlelinecheck=false}
    \caption{\justifying (a) Modulation amplitude in DAMA/LIBRA given the best fit for DAMA/LIBRA and LZ only (green) and for all experiments (blue) in the LO case. The boxes present DAMA's $1\sigma$ confidence intervals and the line the best fit obtained. (b) Posterior of the model parameters for the DAMA/LIBRA and LZ only scenario. The different shades represent the 1, 2, and $3\sigma$ intervals of the combined histograms.}
\end{figure*}

Based on these data, we determine the best fit by applying the MCMC approach described in the previous section. Using these results, we provide updates on past findings, such as \cite{catenaDAMAConfrontsNull2016, kangProbingDAMALIBRA2019}, and explore new combinations of operators. These combinations are shown in table \ref{tab:characterization_operators}. The works in \cite{catenaDAMAConfrontsNull2016, kangProbingDAMALIBRA2019} explore the full parameter space and obtain the results of least tension. They demonstrate how the tension would be alleviated by removing specific experimental results from the analysis or by assuming inelastic scattering. This results in tension of the order $1.5\sigma$, when LUX or PICO is excluded \cite{catenaDAMAConfrontsNull2016}. On the other hand, employing more recent data and inelastic scattering measurements, the tension is reported to be $2.9\sigma$ \cite{kangProbingDAMALIBRA2019}. We analyze the tension between DAMA/LIBRA and the latest results from the other direct detection experiments, most notably LZ.

\subsection{Leading order operators}

Figures \ref{fig:dama_lz_mod_amp} and \ref{fig:dama_lz_posterior} present the modulation amplitude and posteriors obtained from the best fit analysis using solely data from DAMA and LZ, and the LO operators $\OPER_1$ and $\OPER_4$. The corresponding expected number of events in the LZ ROI are $N_\mathrm{LZ} = 342$ events, in stark contrast with the background only result obtained by LZ. This results in a reduced chi-squared value of $\chi^2_\nu = 144.12$. Compared to results shown in \cite{catenaDAMAConfrontsNull2016}, we find that the tension between DAMA/LIBRA and LZ is much stronger with the improved exposure of the LZ experiment. We find a similar tension between DAMA/LIBRA and PICO-60 C$_3$F$_8$. However, although this particular set of operators does not lead to any compatibility, it is possible to construct both the NLO and N$^2$LO cases such that they agree with both experiments simultaneously.

Considering the posterior in figure \ref{fig:dama_lz_posterior}, we observe that the Wilson coefficients $c_1^0$ and $c_1^1$ are centered around zero. Furthermore, these two coefficients do not have any significant correlation with the mass or with the Wilson coefficients of operator $\OPER_4$. In fact, the coefficients contribute only subdominantly to the best fit solution. Consequently, the best fit is driven solely by 
$\OPER_4$ and the mass, with some correlation observed between the mass and each Wilson coefficient. In order to reduce clutter in the following figures, we leave out the posteriors for parameters that do not contribute to the best fit, such as the $c_1$ coefficient in this case.

Our global analysis consists of DAMA, LZ, PICO C$_3$F$_8$ (further PICO Fluorine), PICO CF$_3$I (PICO Iodine) and CDMSlite. The inclusion of PICO Iodine is motivated by \cite{kangProbingDAMALIBRA2019}, which highlights the importance of iodine in constraining the inelastic scattering processes. In PICO Fluorine we neglect carbon as a target, and only consider recoils off the Fluorine nuclei. In addition to the global analysis, which includes all parameters discussed above, we have performed MCMC scans restricted to various subspaces, where only some of the modifications are included, to examine their effects on the tension independently. The simplest case is the EFT analysis without any additional parameters, from here on referred to as \textit{plain}. Secondly, we have cases where this parameter space is extended separately with one of the following: the rescaled velocity dispersion, energy dependent quenching factors, and the inelastic mass splitting parameter. Finally, we also performed a full analysis that includes all of the above parameters.

\begin{figure}
\centering
\includegraphics[width=\linewidth]{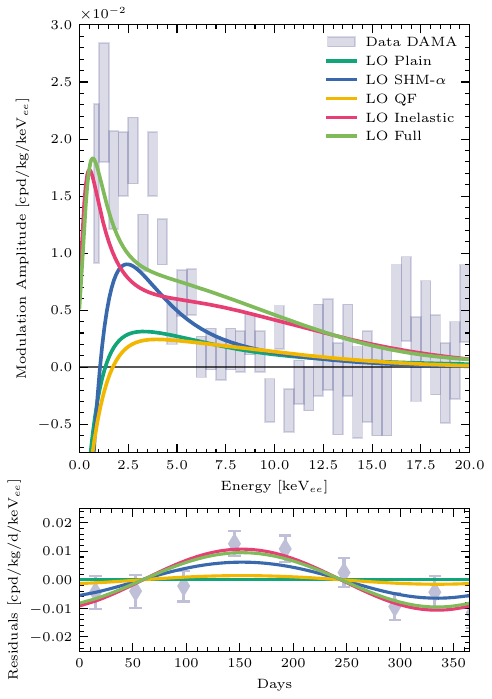}
\captionsetup{justification=RaggedRight,singlelinecheck=false}
\caption{\justifying The upper figure follows the same conventions as Fig. \ref{fig:dama_lz_mod_amp}, showing the best fit for the plain DM non-relativistic LO EFT treatment in green, the modified SHM model in blue, quenching factor model in yellow, inelastic scattering in red, and the full treatment combining all models in light green. The DAMA/LIBRA $1\sigma$ confidence intervals are shown as light blue boxes. Below, we display the annual modulating residual observable in DAMA/LIBRA. The fit is performed over the full time series data released by DAMA/LIBRA, of which we here show an arbitrarily selected one year window. The color scheme for each model follows the legend of the upper figure.}
\label{fig:dama_comparison_LO_all}

\vspace{1em}

\begin{tabular}{l c c c c}
    \toprule
    Case & LZ & PICO Fluorine & PICO Iodine & Total $\chi^2_\nu$\\
    \midrule
    Expected & 18 & 3 & 0 & - \\
    Plain & 207 & 9527 & 409 & $2.25\times10^5$ \\
    SHM-$\alpha$ & 183 & 7914 & 350 & $1.56\times10^5$\\
    QF & 277 & 711 & 8283 & $1.78\times10^5$ \\
    Inelastic & 557 & 0 & 1599 & $1.92\times10^4$ \\
    Full & 529 & 0 & 1993 & $2.97\times10^4$ \\
    \bottomrule
\end{tabular}

\captionsetup[table]{justification=RaggedRight,singlelinecheck=false}
\captionof{table}{\justifying Expected amount of events in LZ, PICO Fluorine (C$_3$F$_8$), PICO Iodine (CF$_3$I), and the corresponding reduced chi-squared value of the global fit. This table lists the values for the LO case described in section \ref{sec:NREFT}.}
\label{tab:LO_best_fit}
\end{figure}

The results of the above analyses are displayed in figure \ref{fig:dama_comparison_LO_all}. The corresponding reduced chi-squared values, in addition to the expected events in each experiment for the best fit parameters, can be found in table \ref{tab:LO_best_fit}. These results are based on the Skellam likelihood function discussed in section \ref{sec:global_analysis}. As discussed earlier, this approach inflates the importance of DAMA/LIBRA with respect to other experiments. Hence, we observe a bias towards fitting the DAMA/LIBRA signal at the cost of massive numbers of expected events in the other experiments. We find that the inclusion of the modified velocity dispersion or the energy dependent quenching factors do not significantly reduce this tension. The inelastic scattering parameter does lead to a decrease of the $\chi^2$ value by an order of magnitude, but the discrepancy still remains statistically extreme. The effect of the inelastic scattering is to suppress the event rate in particular nuclei, due to the modified kinematics. This phenomenon is elaborated on in appendix \ref{sec:min_vel_inelastic}.

We note that the reduced $\chi^2$ value for the inelastic scenario is better than for the global scenario, even though the former is contained in the latter. This is because the Skellam likelihood function which is used in the MCMC analysis is not equal to the Gaussian, which would result in a minimal $\chi^2$. Therefore, the best likelihood in terms of the Skellam distribution does not always coincide with the smallest $\chi^2$, the difference being due to the different effective weight of the DAMA/LIBRA data points in the two approaches. The posterior for the full analysis is shown in appendix \ref{sec:posterior_appendix}. We again observe that the coupling is via $c_4$, while $c_1$ is compatible with zero and is therefore not included in the figure.

As discussed in section \ref{sec:global_analysis}, we have performed an alternative analysis using a Gaussian likelihood function instead of the Skellam distribution. This allows inclusion of the uncertainties of the DAMA/LIBRA modulation amplitudes in the likelihood. Repeating the above analyses with the Gaussian likelihood, we find a best fit where the DAMA/LIBRA signal is suppressed by several orders of magnitude, while the expected number of events in the other experiments are at most $O(10^{-1})$ events. This trend emerged for each set of operators and each model fit we discussed in this work. The reduced chi-squared $\chi^2_\nu$ measure for the Gaussian best fit is significantly better than for the Skellam treatment. We find $\chi^2_\nu = 9.45$, which corresponds to $\approx14\sigma$ deviation to the reported DAMA/LIBRA signal. Similarly, conducting the same analyses for both NLO and N$^2$LO we find approximately the same reduced chi-squared and sigma deviations as reported here. In conclusion, the Gaussian likelihood results in a best fit where the DM signal in every experiment, including DAMA/LIBRA, is essentially zero, and the DAMA/LIBRA result is a statistical fluctuation at the level of $\sim 14\sigma$.

\subsection{Higher order operators}

As demonstrated above, the tensions between DAMA/LIBRA and other null experiments cannot be alleviated with the LO operators, even when allowing for variations in the other model parameters. The higher order operator sets (NLO and N$^2$LO) both enable greater freedom, including a larger number of tunable Wilson coefficients and, specifically for the NLO case, interference terms between different operators. In particular, we adopt the interference terms between $\OPER_8$ and $\OPER_9$ as well as between $\OPER_{11}$ and $\OPER_{12}$. Other interference terms dot not arise in our classification.

\begin{figure}[tbp]
    \centering

    \includegraphics[width=\linewidth]{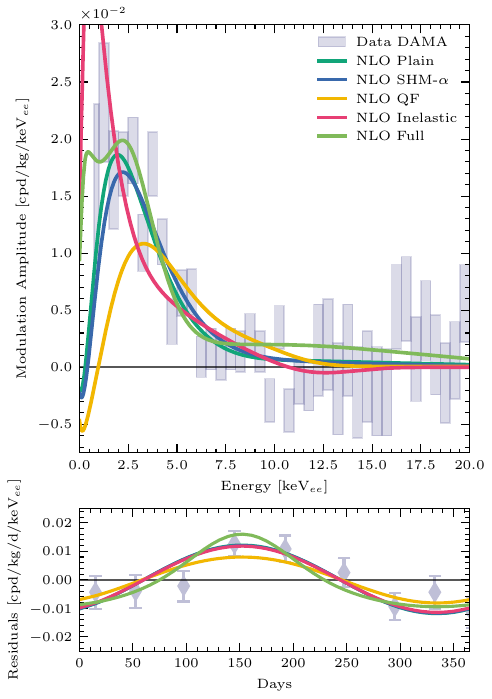}
    \captionsetup{justification=RaggedRight,singlelinecheck=false}
    \caption{\justifying This figure uses the same conventions as figure
    \ref{fig:dama_comparison_LO_all}. Here, we present the results
    for the NLO scenario shown in table \ref{tab:characterization_operators}.}
    \label{fig:dama_comparison_NLO_all}

    \vspace{1em}

    \begin{tabular}{l c c c c}
        \toprule
        Case & LZ & PICO Fluorine & PICO Iodine & Total $\chi^2_\nu$\\
        \midrule
        Expected & 18 & 3 & 0 & - \\
        Plain & 129 & 3221 & 653 & $3.06\times10^4$ \\
        SHM-$\alpha$ & 144 & 1911 & 615 & $1.26\times10^4$\\
        QF & 47 & 532 & 2182 & $3.95\times10^4$ \\
        Inelastic & 453 & 0 & 1161 & $1.1\times10^4$ \\
        Full & 316 & 0 & 877 & $6.41 \times 10^3$\\
        \bottomrule
    \end{tabular}
    \captionsetup[table]{justification=RaggedRight,singlelinecheck=false}
    \captionof{table}{\justifying Expected amount of events in LZ, PICO Fluorine
    (C$_3$F$_8$), PICO Iodine (CF$_3$I) and the corresponding reduced
    chi-squared value of the global fit. This table lists the values
    for the NLO case described in section \ref{sec:NREFT}.}
    \label{tab:NLO_counts}
\end{figure}

\begin{figure}[]
    \centering

    \centering
    \includegraphics[width=\linewidth]{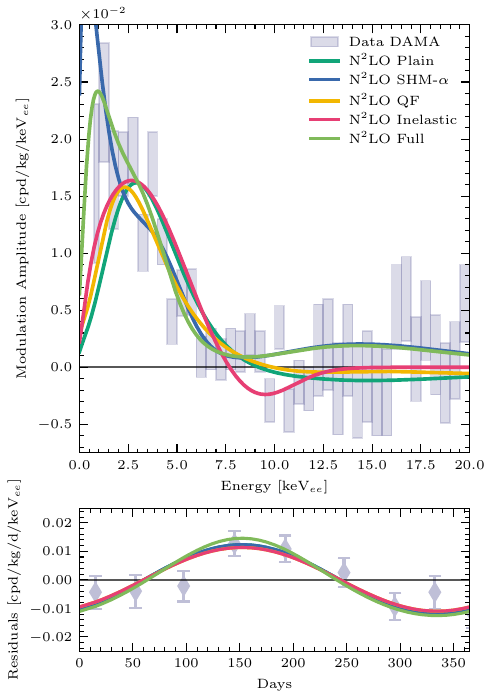}
    \captionsetup{justification=RaggedRight,singlelinecheck=false}
    \caption{\justifying We once more follow the labeling convention used in figures \ref{fig:dama_comparison_LO_all} and \ref{fig:dama_comparison_NLO_all}, and employ the N$^2$LO operators listed in table \ref{tab:characterization_operators}.}
    \label{fig:dama_comparison_NNLO_all}
    \vspace{1em}

    \begin{tabular}{l c c c c}
        \toprule
        Case & LZ & PICO Fluorine & PICO Iodine & Total $\chi^2_\nu$\\
        \midrule
        Expected & 18 & 3 & 0 & - \\
         Plain & 41 & 1192 & 2062 & $2.31\times10^4$\\
         SHM-$\alpha$ & 72 & 766 & 904 & $7.91\times10^3$\\
         QF & 58 & 1727 & 1258 & $2.85\times10^4$\\
         Inelastic & 127 & 2 & 1537 & $1.84\times10^4$\\
         Full & 50 & 301 & 874 & $6.46\times10^3$\\
         %Gaussian & - & - & - & -\\
         \bottomrule
    \end{tabular}
    \captionsetup[table]{justification=RaggedRight,singlelinecheck=false}
    \captionof{table}{\justifying Expected amount of events in LZ, PICO Fluorine (C$_3$F$_8$), PICO Iodine (CF$_3$I) and the corresponding reduced chi-squared value of the global fit. This table lists the values for the N$^2$LO case described in section \ref{sec:NREFT}.}
    \label{tab:NNLO_counts}
\end{figure}

We repeat the MCMC treatment applied to the LO case for the NLO and N$^2$LO operators, using the Skellam and Gaussian likelihood functions. The latter yields again a $\approx14\sigma$ deviation for DAMA/LIBRA with the modulation amplitude suppressed by several orders of magnitude. The analyses using the Skellam distribution for the NLO operators and N$^2$LO operators are presented in figures \ref{fig:dama_comparison_NLO_all} and \ref{fig:dama_comparison_NNLO_all}. The corresponding expected counts in the other experiments are listed in tables \ref{tab:NLO_counts} and \ref{tab:NNLO_counts}.

The NLO case follows a similar trend as the LO case; that is, the inclusion of the mass splitting parameter yields the most significant improvement of the three models. In this case, the full model space performs better in terms of the reduced $\chi^2$ than the inelastic model. The posterior of the NLO case, illustrated in appendix \ref{sec:posterior_appendix}, shows that for this set of operators the best fit DM velocity dispersion scaling parameter is approximately $\alpha \approx 0.11$ in comparison to $\alpha \approx 1$ for the LO case. Thus, for the NLO model a narrower DM velocity distribution is preferred over the SHM, as this increases the relative annual modulation amplitude. Moreover, the quenching factors obtained from this best fit analysis remain constant over the relevant energy range, with $\text{QF}_{\mathrm{Na}} = 0.025$ and $\text{QF}_{\mathrm{I}} = 0.089$.

The drastic change in quenching factor arises because of the inelastic scattering kinematics in this best fit model. Sodium recoils are only kinematically allowed above approximately $\sim\unit[19]{keV_\mathrm{nr}}$. The small quenching factor thus compensates for the high nuclear recoil energies, so that the recoil spectrum in DAMA/LIBRA appears at the correct electron equivalent energy range. This scenario demonstrates the non-trivial interplay of our global model parameters that can lead to an improved global fit. Regardless, it is not sufficient to remove the tension.

The N$^2$LO results exhibit an overall behavior similar to the NLO results, although some differences arise in the relative performance of the individual model treatments. In this case, the modified halo model provides the best fit among all single-model extensions. As in the NLO case, here, too, the global model leads to the best fit result in terms of the reduced $\chi^2$, indicating that the additional freedom provided by the combined treatment is significant for attempting to accommodate the constraints from the different null experiments. The posterior for the N$^2$LO case, shown as the last figure in appendix \ref{sec:posterior_appendix}, similarly favors a reduced velocity dispersion, with a best-fit dispersion scaling factor of $\alpha \approx 0.38$. The quenching factors also follow the same qualitative trend as observed for the NLO operators. We find $\text{QF}{\mathrm{Na}} = 0.012$ and $\text{QF}{\mathrm{I}} = 0.089$. In particular, the iodine quenching factor remains essentially unchanged, while the sodium quenching factor is reduced even further compared to the NLO case. This is due to the same interplay between the inelastic kinematics as discussed above for the NLO case. 

Neither the NLO nor the N$^2$LO case qualitatively alters the conclusion drawn from the LO analysis. Among the minima identified using our MCMC methodology, the NLO case provides the best overall fit. The Skellam and Gaussian likelihood treatments yield substantially different levels of statistical exclusion, but nevertheless lead to the same qualitative conclusion. Namely, within the assumptions and analysis framework adopted in this work, the DAMA/LIBRA modulation remains incompatible with the constraints set by other null experiments such as LZ, PICO-60 and CDMSlite. 

\section{Conclusion}
\label{sec:checkout}

We have investigated the compatibility of the annual modulation reported by DAMA/LIBRA with the results of other direct detection experiments, considering LZ, CDMSlite, PICO-60 C$_3$F$_8$ and PICO-60 CF$_3$I. To enable a broad range of possible DM scattering scenarios, we performed a global analysis using a non-relativistic EFT description of the DM--nucleon interaction. Additionally, we introduced extensions of the standard assumptions concerning the DM velocity distribution, the NaI quenching factors, and inelastic scattering. The resulting parameter space was explored using MCMC methods, allowing us to investigate both the quality of the global fits and correlations between the different model parameters.

Our results indicate that the additional freedom within the DM--nucleon scattering realm can substantially modify how the DAMA/LIBRA modulation is obtained relative to the null experiments. In particular, of the extended models, inelastic scattering provides the largest reduction in its tension with other experiments. The higher-order EFT operators introduce additional degrees of freedom and interference terms, but do not qualitatively change the conclusions obtained at leading order: the best fit solution identified in the NLO analysis remains statistically incompatible with the null experiments. We notice that parameter values which suppress the predicted rates in xenon and fluorine instead result in larger rates in iodine, illustrating the difficulty of simultaneously satisfying the constraints of the different target materials.

We performed our MCMC analysis using two different choices for the DAMA/LIBRA likelihood function. Using a Skellam distributed likelihood favors a good fit to the DAMA/LIBRA modulation data, at the cost of larger event rates in the other experiments. These results can be interpreted as the best one can do to minimize the rates in the other experiments, while keeping the DAMA/LIBRA modulation fit reasonably good. In contrast, treating the published DAMA/LIBRA modulation amplitudes as Gaussian distributed with their reported uncertainties results in a best fit that strongly suppresses the fitted modulation amplitude. We find a deviation of approximately $14\sigma$ in this treatment, comparable with the significance of the DAMA/LIBRA observation of a non-zero modulation amplitude. The result is the same across the LO, NLO, and N$^2$LO operator sets. This implies that the combined statistical significance of the non-observation of DM in the other experiments is still much stronger than the significance of the positive observation in DAMA/LIBRA. Since a compromise that fits both cannot be found, as was seen in the Skellam-analysis discussed above, the combined analysis then results in essentially zero global DM signal for the Gaussian likelihood.

In summary, although the numerical significance depends strongly on the statistical treatment of the DAMA/LIBRA data, the two likelihood constructions lead to the same qualitative conclusion: the DAMA/LIBRA signal is incompatible with the non-observation of DM in the other experiments, and this conclusion remains even in the extended model space we have analyzed.

\begin{acknowledgments}

This work has been supported by the Research Council of Finland (grants$\#$ 342777, 371542). Further, it has been supported by the  Vilho, Yrjö and Kalle Väisälä fund and the Finnish Graduate School in Particle and Nuclear Physics (Doctoral Education Pilot). The authors wish to thank IT for Science at the University of Helsinki for supporting this project with computational and datastorage resources. HILE is funded by ERC grant(ILLUMINATOR, 101114623).

\end{acknowledgments}

\appendix
\section{Minimum velocity for inelastic scattering}
\label{sec:min_vel_inelastic}
\begin{figure}
    \centering
    \captionsetup{justification=RaggedRight,singlelinecheck=false}
    \includegraphics[width=\linewidth]{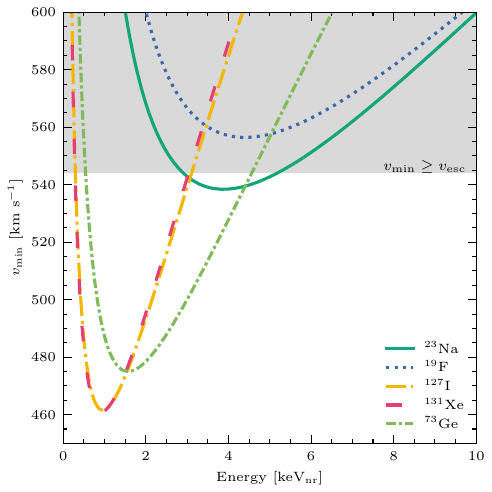} % , trim=0 0 0 -10
    \caption{\justifying Minimum DM velocity to generate a nuclear recoil of a given energy for nuclei relevant in this work, with parameters $\delta=\unit[12]{keV}$ and $m_\chi = \unit[11]{GeV}$. The region where the required velocity exceeds the maximum velocity of DM particles in the SHM is shown in grey.}
    \label{fig:exclusion_vmin}
\end{figure}
The minimum DM velocity in the laboratory frame required to excite a nuclear recoil energy $E_{\rm nr}$ is given by equation \eqref{eq:vmin}, including the mass splitting parameter $\delta$. In the case of elastic scattering, $\delta =0$, the minimum velocity is proportional to $\sqrt{E_\mathrm{nr}}$, a strictly increasing function of recoil energy. In the inelastic scattering case, the mass splitting $\delta$ introduces a global minimum in $v_\text{min}$ at $E_\text{nr} = \delta \mu/m_T$, where $v_\text{min} = \sqrt{2\delta/\mu}$. Therefore, for some combinations of mass splitting and reduced mass, $v_\text{min}$ can exceed the escape velocity at all energies. For sufficiently massive DM such that the reduced mass $\mu$ is significantly impacted by the mass of the target nucleus $m_T$, whether this limit is exceeded can vary drastically for different nuclei. Consequently, the scattering off specific nuclei can become kinematically forbidden, while scattering off others is allowed. In a less extreme scenario, the scattering rate on some targets can be significantly reduced due to the suppression of the available phase space. Figure \ref{fig:exclusion_vmin} illustrates the effect for various nuclei, where for the given set of DM mass splitting and DM mass, scattering off iodine, xenon, and germanium is allowed in a wide energy band, while scattering off sodium is restricted to the high energy tail of the DM velocity distribution, and scattering off fluorine is forbidden.

\section{Posterior distributions based on the full treatment}
\label{sec:posterior_appendix}
Below we show the posterior distributions in the global analysis for the LO, NLO and N$^2$LO operators.

\begin{figure*}
    \centering
    \includegraphics[width=\textwidth]{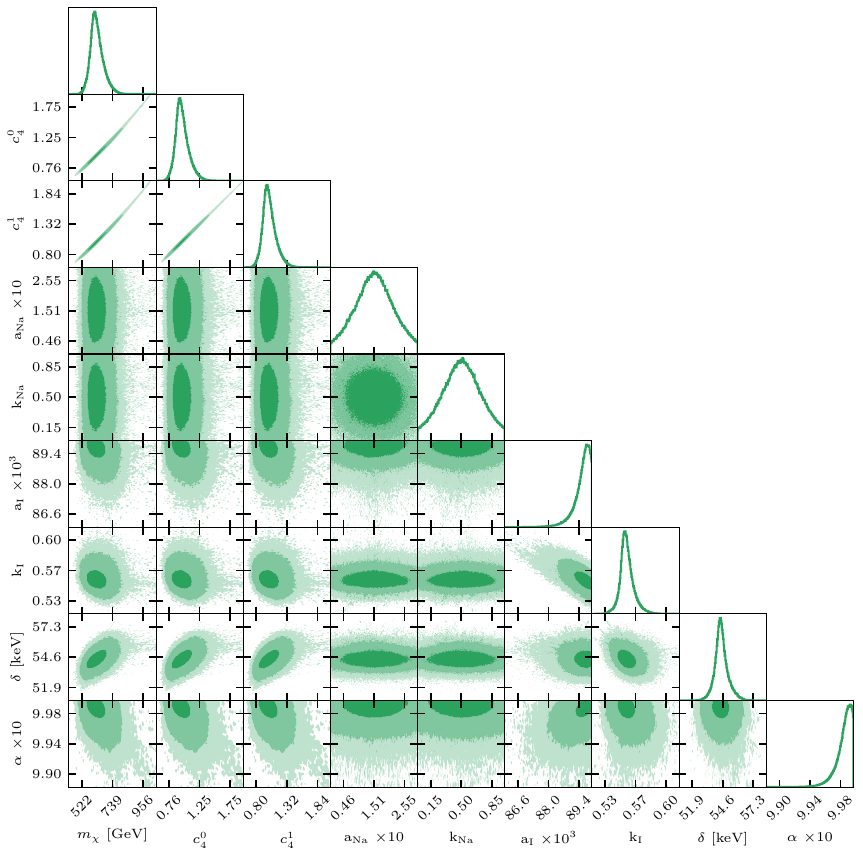}
    \captionsetup{justification=RaggedRight,singlelinecheck=false}
    \caption{\justifying The posterior distributions for the global analysis with LO operators. The DM mass is given in GeV, the Wilson coefficients, given the relativistic normalization, are unitless and the mass splitting parameter $\delta$ in keV. The different shaded regions portray the 1, 2 and 3$\sigma$ bounds of our best fit, from dark to lighter green.}
    \label{fig:LO_entire_posterior}
\end{figure*}

\begin{figure*}
    \centering
    \includegraphics[width=\textwidth]{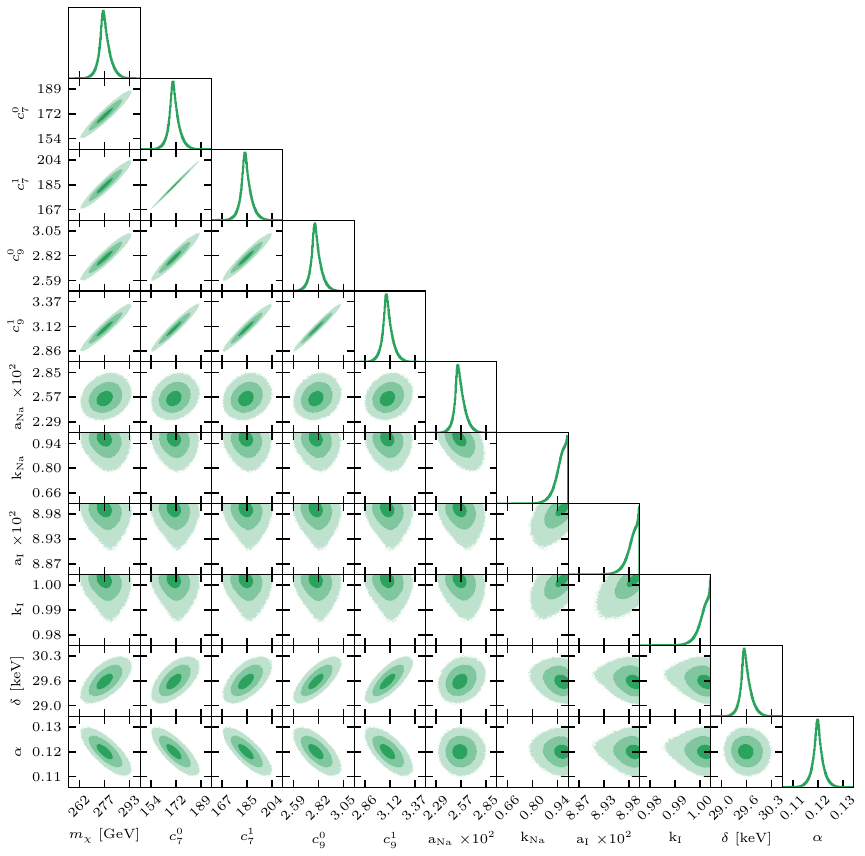}
    \captionsetup{justification=RaggedRight,singlelinecheck=false}
    \caption{\justifying The posterior distributions for the global analysis with NLO operators. The DM mass is given in GeV, the Wilson coefficients, given the relativistic normalization, are unitless and the mass splitting parameter $\delta$ in keV. The different shaded regions portray the 1, 2 and 3$\sigma$ bounds of our best fit, from dark to lighter green.}
    \label{fig:NLO_entire_posterior}
\end{figure*}

\begin{figure*}
    \centering
    \includegraphics[width=\textwidth]{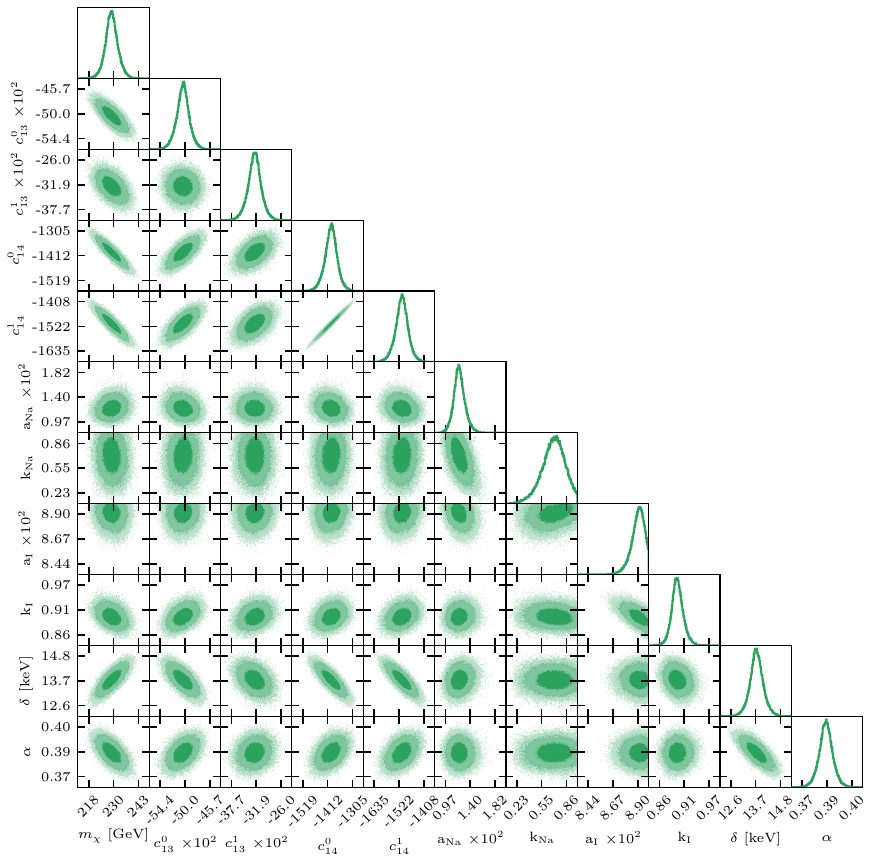}
    \captionsetup{justification=RaggedRight,singlelinecheck=false}
    \caption{\justifying The posterior distributions for the global analysis with N$^2$LO operators. The DM mass is given in GeV, the Wilson coefficients, given the relativistic normalization, are unitless and the mass splitting parameter $\delta$ in keV. The different shaded regions portray the 1, 2 and 3$\sigma$ bounds of our best fit, from dark to lighter green.}
    \label{fig:NNLO_entire_posterior}
\end{figure*}

\clearpage

%%%%%%%%%%%%%%%%%%%%%%%%%%%%%%%%%%%%%%%%%%%%%%%%
\bibliography{refs.bib}
%%%%%%%%%%%%%%%%%%%%%%%%%%%%%%%%%%%%%%%%%%%%%%%%%%

%%%%%%%%%%%%%%%%%%%%%%%%%%%%%%%%%%%%%%%%%%%%%%%%
%%%%%%%%%%%%%%%%%%%%%%%%%%%%%%%%%%%%%%%%%%%%%%%%

\end{document}